\documentclass[%
reprint,
superscriptaddress,
amsmath,amssymb,
aps,
prl,
floatfix
]{revtex4-2}
\usepackage{graphicx} % Required for inserting images
\usepackage{mathtools}
\usepackage{braket}
\usepackage{booktabs}
\usepackage{xcolor}
\usepackage{float}
\usepackage{orcidlink}
\usepackage{multirow}
\usepackage{adjustbox}

\newcommand{\mat}[1]{\mathcal{#1}}
\newcommand{\ham}{\hat{\mathcal{H}}}
\renewcommand{\ket}[1]{\lvert #1 \rangle}
\renewcommand{\bra}[1]{\langle #1 \rvert}
\renewcommand{\vec}[1]{\boldsymbol{#1}}

\newcommand{\prlsec}[1]{\textit{#1.}---}

\usepackage[capitalise]{cleveref}
\begin{document}
\title{Quantum-induced latent symmetries}
\date{6 November 2026}
\author{Jonas Himmel\orcidlink{0009-0007-2943-5197}}
\affiliation{Institute of Physics, University of Rostock, Rostock, Germany}
\author{Malte Röntgen\orcidlink{0000-0001-7784-8104}}
\email{mroentgen@eitech.edu.cn}
\affiliation{Eastern Institute for Advanced Study, Eastern Institute of Technology, Ningbo, China}
\author{Tilman G\"uhlstorf}
\affiliation{Institute of Physics, University of Rostock, Rostock, Germany}
\author{Matthias Heinrich\orcidlink{0000-0003-3552-0789}}
\affiliation{Institute of Physics, University of Rostock, Rostock, Germany}
\author{Max Ehrhardt\orcidlink{0000-0001-8653-1168}}
\affiliation{Institute of Physics, University of Rostock, Rostock, Germany}
\author{Alexander Szameit\orcidlink{0000-0003-0071-6941}}
\email{alexander.szameit@uni-rostock.de}
\affiliation{Institute of Physics, University of Rostock, Rostock, Germany}
\author{Tom A. W. Wolterink\orcidlink{0000-0002-3591-8044}}
\affiliation{Independent researcher, The Netherlands}
\begin{abstract}
We study the interplay of latent symmetries and multiparticle quantum states upon the introduction of additional particles. We show how physical symmetries in multimode quantum systems give rise to symmetries in the multiparticle state space of indistinguishable fermions and bosons.
Latent symmetries, hidden symmetries that are exclusively visible in the eigenspectral domain, offer new ways to engineer the functionality of multimode quantum systems. In this vein, we present multiboson state generation as an application of latent symmetries to the multiparticle domain. Finally, we demonstrate that even in systems that are entirely devoid of symmetry, latent symmetries can be induced in the multiparticle state space, as a result of quantum indistinguishability.
\end{abstract}

\maketitle

\prlsec{Introduction}%
The nonclassical behaviour of indistinguishable particles is one of the most remarkable features of quantum physics. A key aspect of quantum indistinguishability is that different particle types obey different exchange statistics, which lead to fundamentally different behaviour: fermions obey the Pauli exclusion principle, whereas bosons can bunch together in the same state. In large-scale quantum systems, where many particles jointly occupy many modes, quantum interference between indistinguishable pathways constitutes a valuable resource for applications in quantum computing, metrology, and quantum communication \cite{nielsenQuantumComputationQuantum2010}.

Symmetries of the underlying system strongly influence quantum interference and help to engineer systems with desired functionalities.
While they may be apparent at first glance, in the case of conventional spatial symmetries, symmetries can also be latent, i.e.\ hidden in the system's eigenspectrum. The concept of latent symmetries originates from graph theory \cite{smithHiddenSymmetriesReal2019,kemptonCharacterizingCospectralVertices2020,godsilStronglyCospectralVertices2024}, and has since helped to understand phenomena such as degeneracies \cite{rontgenLatentSymmetryInduced2021}, flat bands \cite{morfoniosFlatBandsLatent2021}, topology \cite{eekHigherorderTopologyProtected2025,eekEmergentNonHermitianModels2024,zhangLatentGeometryCorrespondenceUnraveling2026} and state transfer \cite{eisenbergPrettyGoodQuantum2019, rontgenDesigningPrettyGood2020, vanBommel2019QIC190601CompleteCharacterizationPrettyGood, himmel2026}.
So far, latent symmetries have been treated as intrinsic properties of the physical system, separately from the quantum nature of the particles involved \cite{himmel2025}. It remains an open question how quantum indistinguishability affects these symmetries, and whether it can induce or suppress latent symmetries in multiparticle systems.

In this work, we show how quantum indistinguishability shapes the latent symmetries of multiparticle quantum systems. Using graph-theoretical methods, we determine how single-particle symmetries are inherited by systems of non-interacting bosons and fermions. 
We demonstrate the versatility of this framework by applying it to the generation of multiphoton states. 
Beyond mere inheritance, we reveal that quantum indistinguishability by itself can establish multiparticle latent symmetries, even when the underlying single-particle system intrinsically possesses neither spatial nor latent symmetry.

\begin{figure}[!htt]
    \includegraphics[width=\columnwidth]{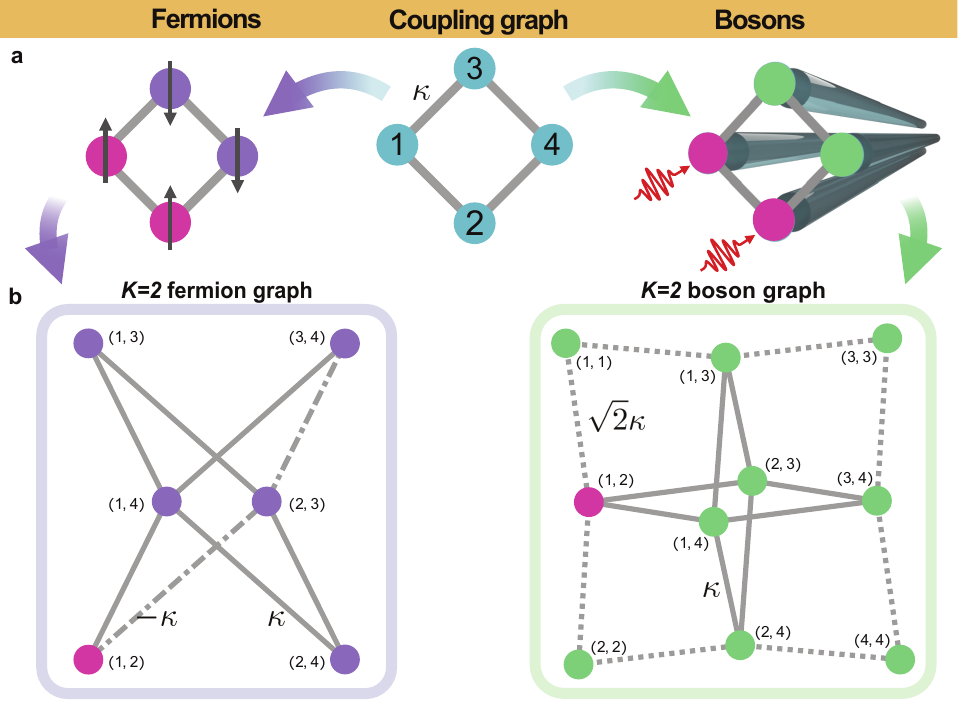}
    \caption{\label{fig:fockGraphIntro}
    Multiparticle graphs. \textbf{a} Any coupled-mode quantum system, such as fermionic coupled spin networks or coupled photonic waveguides, can be represented by a coupling graph. The behaviour of a single particle can be described as a quantum walk on this graph. For an excitation of $K=2$ indistinguishable particles, one has to take into account their fermionic or bosonic nature as well. \textbf{b} Using a multiparticle graph representation, where sites indicate basis states, the resulting dynamics can be described by a quantum walk on the multiparticle graph. Here, site (1,2) e.g. corresponds to the basis state $\ket{\vec{m}}$ of one particle being in mode 1 and the other in mode 2.}
\end{figure}

\prlsec{Indistinguishable particles on graphs}%
Systems containing $N$ coupled modes can be mapped onto an $N$-site coupling graph that encodes the physical structure of the system, where each site corresponds to a single mode and the edges are the couplings between them (Fig.~\ref{fig:fockGraphIntro}a).
Throughout this work, we investigate the behaviour of such coupled-mode networks when populating them with non-interacting indistinguishable bosons or fermions.
The dynamics of a single particle in such a network are independent of the particle's nature and therefore fully described by a quantum walk on the couling graph.
In contrast, for $K$ indistinguishable particles, the second-quantized Hamiltonian $\ham=\sum_{ij}H_{ij}\hat{a}_i^\dagger \hat{a}_j$, with $\hat{a}_i^\dagger$ ($\hat{a}_i$) the creation (annihilation) operators, gives rise to distinct behaviour of fermions and bosons, according to their different exchange statistics as determined by their respective commutation and anti-commutation relations. One way to capture this multiparticle behavior is to consider the time evolution of the coupled-mode system first and subsequently add information on the exchange statistics \cite{tichyManyparticleInterferenceManyboson2012}. 
Here, we express the Hamiltonian as a graph of size $\binom{N}{K}$ for fermions and $\binom{N+K-1}{K}$ for bosons: its sites represent the basis states $\ket{\vec{m}}$, which count how many particles occupy each mode, and edges between two sites $\ket{\vec{m}}$ and $\ket{\vec{n}}$ are the matrix elements $\langle\vec{m}|\ham|\vec{n}\rangle$ (see Fig.~\ref{fig:fockGraphIntro}b for an example for $K=2$).
The full evolution of $K$ particles across $N$ modes of the system is therefore determined by the time evolution of an excitation of the multiparticle graph, intrinsically containing the particle nature \cite{peruzzo2010,tschernig2020,himmel2025}. This enables the application of graph-theoretical tools for multiparticle dynamics.
For a more detailed description of the construction of these multiparticle graphs, see End Matter. 

\begin{figure*}[!htb]
    \includegraphics[width=0.9\textwidth]{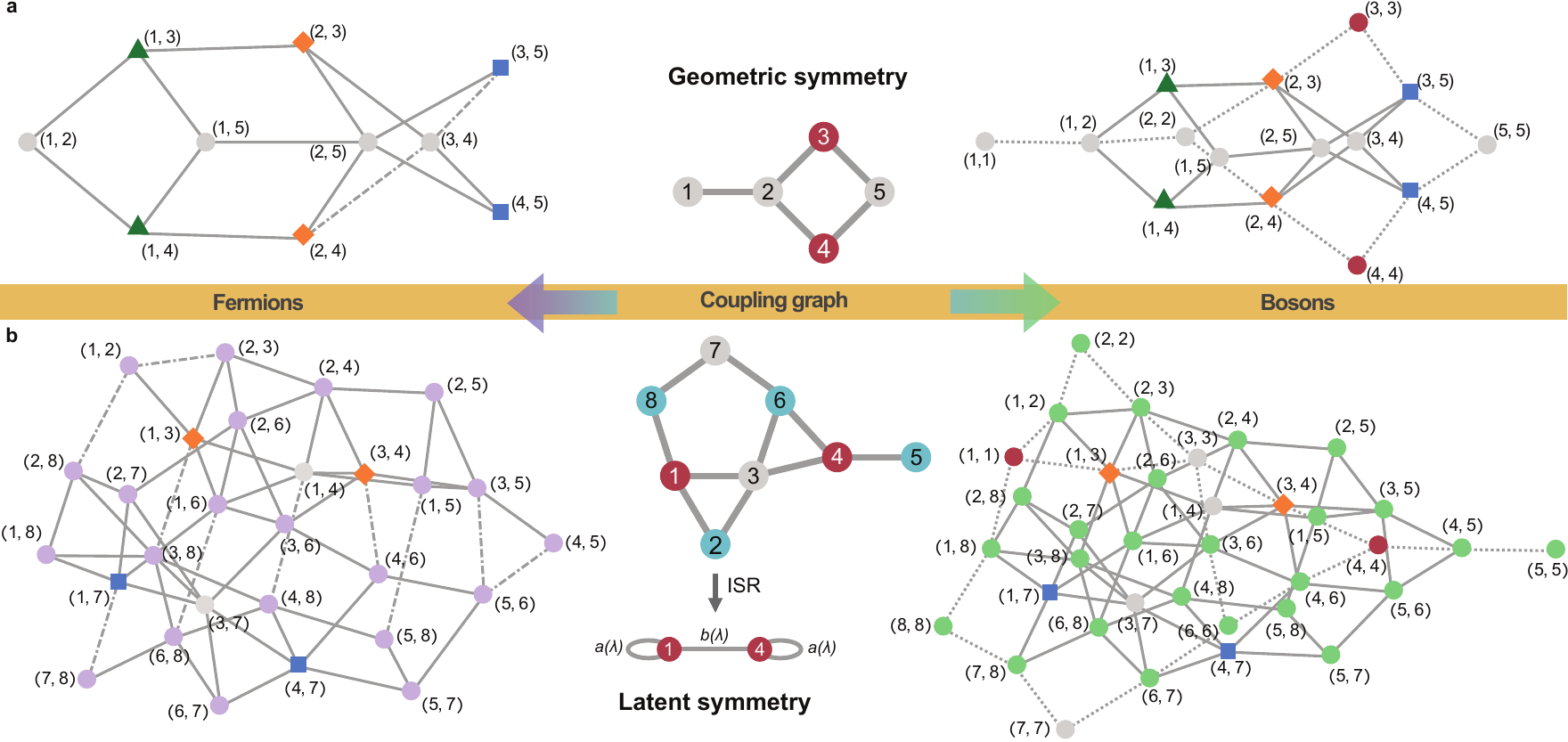}
    \caption{\label{fig:fockGraphSymm}
    Symmetries in Fock graphs. \textbf{a} An example graph featuring a real-space permutation symmetry between sites $u=3$ \& $v=4$ (red) and singlet sites $s_1=1$, $s_2=2$ \& $s_3=5$ (grey). The symmetry is transferred to the fermionic and bosonic $K=2$ multiparticle graphs. Two-particle sites corresponding to one particle at one of the symmetric sites $u,v$ and the other at a singlet site $s_i$ become symmetric (green triangle, orange diamond, blue square). Two-particle sites corresponding to both particles on the same site (red circle) become symmetric for bosons and vanish for fermions. Both particles each at a different symmetric site becomes a two-particle singlet site, here (3,4). \textbf{b} A latent symmetric graph whose isospectral reduction over the two sites $u=1$ \& $v=4$ (red) yields a symmetric graph. Sites $s_1=3$ \& $s_2=7$ are singlet sites (grey). Constructing the fermionic and bosonic multiparticle graphs shows the same transfer pattern as for permutation symmetries shown in \textbf{a}.}
\end{figure*}

\begin{figure}[hb!]
    \centering
    \includegraphics[width=\linewidth]{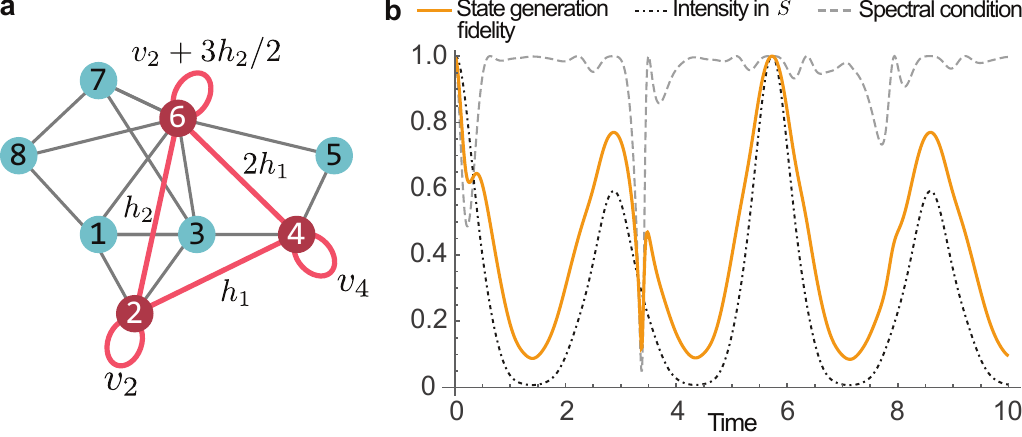}
    \caption{
   \textbf{a}
    A single-particle graph with ($\alpha=2$) scaled latent symmetry for $u=2$, $v=6$, and $s=4$ a singlet, which is preserved by changing the couplings $h_{1,2}$ and on-site potentials $v_{2,4}$.
     \textbf{b} The time evolution of the two-photon state $(2,4)$ shows fractional revival at $t_\mathrm{F} \approx 5.72$. The dashed curves show the total intensity in $S=(\vec{u},\vec{v})$ and the fulfilment of the spectral condition respectively. Only when both are close to 1, the desired state is generated. The state generation fidelity is obtained as the square root of the product of the intensity and spectral condition line. All unlabelled lines correspond to a coupling strength of unity; the expressions for  $h_1,h_2,v_2,v_4$ and for the spectral condition are given in the End Matter.
}
    \label{fig:fractionalRevival}
\end{figure}

\begin{figure*}[!hbt]
    \includegraphics[width=0.9\textwidth]{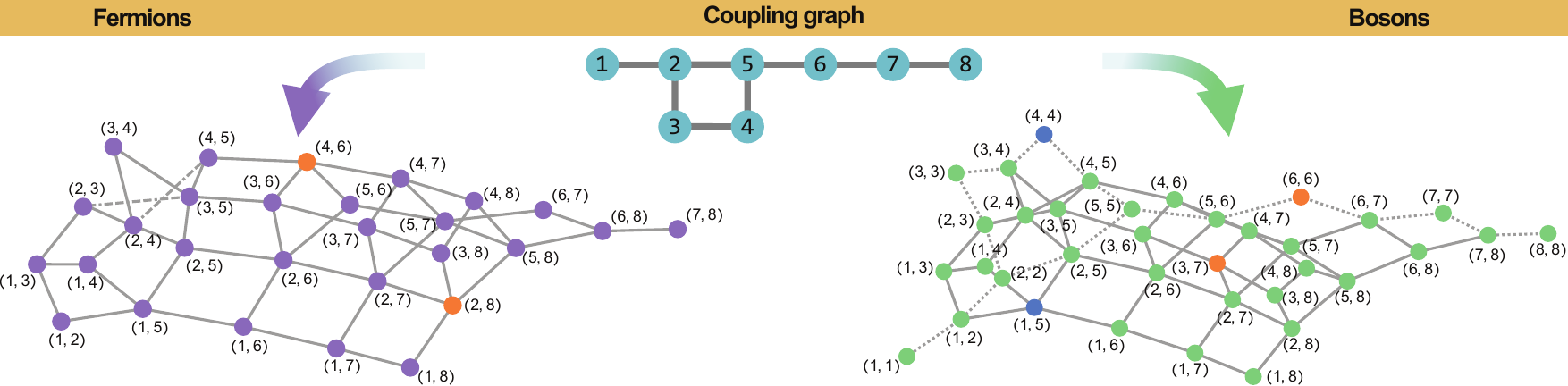}
    \caption{Quantum-induced latent symmetries in a completely asymmetric graph, without any real space or latent symmetries. For two particles, the fermionic and bosonic graphs show the emergence of different latent symmetric site pairs (orange and blue). The fermionic graph features one symmetric pair whereas the bosonic graph contains two.}
    \label{fig:emergSymm}
\end{figure*}
\prlsec{Graph symmetries}
The most prominent examples of symmetries are visible to the naked eye, such as reflections, rotations, inversions, and combinations thereof. They are described by permutation symmetries of the underlying graph: two sites $u,v$ of a graph are permutation symmetric if the permutation matrix $\mat{P}$ that exchanges them commutes with the  graph's adjacency matrix (\cref{fig:fockGraphSymm}a). 
Likewise, symmetries may also reside in the eigenspectral domain while remaining invisible in the physical geometry of the system. These are called latent or hidden symmetries \cite{smithHiddenSymmetriesReal2019,kemptonCharacterizingCospectralVertices2020,godsilStronglyCospectralVertices2024}. They can be detected by performing an isospectral reduction (ISR) \cite{bunimovichIsospectralTransformations2014} over a subset of sites $S$ (see End Matter).
This operation reduces the system's dimensionality while preserving its eigenvalue spectrum.
If the resulting reduced system is symmetric under exchange of two sites $u,v \in S$ [\cref{fig:fockGraphSymm}b]---and hence commutes with a permutation matrix $\mat{P}$---then these sites are latent symmetric.

Latent symmetries can be understood as a generalization of spatial symmetries: In absence of degeneracies \footnote{In the case of degenerate eigenvalues, eigenmodes can be chosen to have definite parity.}, systems featuring either one of the two symmetries fulfil the same condition on all of their eigenmodes $\ket{\lambda_r}$,
\begin{equation}
    \braket{u|\lambda_r}=\pm\braket{v|\lambda_r},
    \label{eq:paritySinglePart1}
\end{equation}
implying definite parity on $u$ and $v$ for all eigenmodes, where $\braket{u|\lambda_r}$ denotes the component of $\ket{\lambda_r}$ on site $u$.

Given a graph with a symmetry, either spatial or latent, it may happen that all negative-parity eigenmodes vanish on certain sites of the graph.
Following the terminology of Ref.~\cite{Morfonios2021LAaiA62453CospectralityPreservingGraphModifications}, we refer to these sites as \emph{singlets}.
For geometric symmetries, those sites are left invariant by the symmetry operation and can be easily found by eye.

\prlsec{Multiparticle symmetry inheritance}%
While symmetries of the coupling graph directly determine the dynamics of single particles, they also influence the multiparticle regime. 
In direct analogy to \cref{eq:paritySinglePart1}, when two sites of a multiparticle graph corresponding to basis states $\ket{\vec{u}},\ket{\vec{v}}$ are latent symmetric, the eigenmodes $\ket{\vec{\Phi}}$ have definite parity on these sites,
\begin{align}
    \braket{\vec{u}|\vec{\Phi}}=\pm\braket{\vec{v}|\vec{\Phi}}.
    \label{eq:multi-partLSMain}
\end{align}

We now study how systems fulfilling \cref{eq:paritySinglePart1} behave in the multiparticle regime and find that spatial and latent symmetries are inherited according to the same pattern (\cref{fig:fockGraphSymm}, $K=2$): Although the original single-particle system only features one symmetric pair, the bosonic and fermionic two-particle graphs now feature more than one symmetric pair. The additional two-particle symmetric pairs generally correspond to one particle on a single-particle singlet and the other particle on one of the single-particle symmetric sites (blue, green, orange). In turn, the presence of both particles at a single-particle singlet yields a singlet site in the two-particle graphs, as does the site corresponding to one particle at each of the single-particle symmetric sites (gray). For bosons, the site corresponding to both particles being on the same single-particle symmetric site is inherited as a symmetric pair (red) that vanishes for fermions due to Pauli exclusion. 
As a consequence, fermionic systems without single-particle singlets generally do not have any inherited latent symmetry for $K=2$; we refer to this as symmetry suppression by indistinguishability.
The above principle can be readily generalized to arbitrary $K$ (see End Matter).

\prlsec{Dynamics and state generation}%
Let us now consider the dynamics of our system, mediated through the time evolution operator $\hat{\mathcal{U}}(t) = exp(-\mathrm{i}\, \ham t)$ acting on some initial state $\ket{\vec{\Psi}_0}$.
As $\ket{\vec{\Psi}_0}$ can be expanded in the eigenmodes, both geometric and latent symmetries leave a clear mark on the dynamics and one can e.g. harness symmetries of a system to facilitate state transfer. By fine-tuning some of the elements, while maintaining the overall symmetry, it is possible to construct a graph that transfers an initial excitation of $\ket{\vec{u}}$ to $\ket{\vec{v}}$ with high or even perfect fidelity by simple time evolution \cite{Bose2003PRL91207901QuantumCommunicationUnmodulatedSpin,Christandl2005PRA71032312PerfectTransferArbitraryStates,Kay2010IJQI08641PerfectEfficientStateTransfer,Kendon2011JoCaTN8422PerfectStateTransferQuantum,himmel2026,Godsil2012PRL109050502NumberTheoreticNatureCommunicationQuantum,Godsil2012DM312129StateTransferGraphs}.

Beyond transferring $\ket{\vec{u}}$ to $\ket{\vec{v}}$, let us broaden our view by realising that the strict parity condition, \cref{eq:paritySinglePart1}, is a special case of a larger class of spectral relations.
It can be relaxed to a scaled $\alpha$ parity condition \cite{Chan2020AQFundamentalsFractionalRevivalGraphs}, where each eigenmode fulfils one of the two relations
\begin{align}
    \label{eq:scaledParitySinglePart}
    %\end{split}
    \braket{u|\lambda_r}=\alpha\braket{v|\lambda_r},\ \braket{u|\lambda_r}=- \frac{1}{\alpha}\braket{v|\lambda_r}.
\end{align}
with a real, non-vanishing $\alpha$, while conventional parity is recovered for $\alpha = 1$.

\Cref{eq:scaledParitySinglePart} is equivalent to the statement that the isospectral reduction over $S=\{u,v\}$ commutes with the matrix $\begin{psmallmatrix}
    0 & 1 \\ 1 & \beta
\end{psmallmatrix}$
with $\beta = \alpha - 1/\alpha$.
Importantly, for $\alpha \ne 1$, this is not a permutation matrix but rather an example of a generalized latent symmetry.
In other words, the underlying graph has a $\alpha$-scaled latent symmetry, which, for $\alpha = 1$, becomes a conventional latent symmetry.
Importantly, this scaled symmetry can likewise be inherited by the multiparticle graphs.
Specifically, if the single-particle graph contains a singlet $s$ \footnote{Which is now defined as the site on which all negative scaled parity eigenmodes vanish; it can be easily shown that this is equivalent to $\alpha (H^k)_{u,s}=(H^k)_{v,s}$ for any $k$.}, then the two-boson states $\ket{\vec{u}} = \ket{(w,u)}$, $\ket{\vec{v}} = \ket{(w,v)}$ are $\alpha$-scaled latently symmetric.

Scaled latent symmetry has two interesting applications.
In waveguide networks, it leads to certain broadband linear relations between scattering coefficients \cite{rontgenGeometricSymmetryBroadband2026}.
In single-particle graphs, one can tune the system to produce fractional revival \cite{chan2022}, as there exists a time $t_F$ such that
\begin{equation}
    \hat{\mathcal{U}}(t_\mathrm{F})\ket{\vec{u}}
    =
    a\ket{\vec{u}}+b\ket{\vec{v}},
    \label{eq:stateGeneration}
\end{equation}
where $|a|^2+|b|^2=1$. The scaled parity relation restricts the relative magnitude and phase of the two amplitudes through
\begin{equation} \label{eq:realPartCond}
    \operatorname{Re}\left(\frac{a}{b}\right)
    =
    -\frac{\beta}{2} = - \frac{\alpha - 1/\alpha}{2}.
\end{equation}
Thus, the amplitudes $a$ and $b$ cannot be chosen independently. Nevertheless, fractional revival provides a means of generating superpositions and, if the two states describe distinct multiparticle configurations, multiparticle entanglement.

Here, we propose to use this idea as a mechanism for multiphoton state generation. The central challenge is that the sites and edges of a multiparticle graph are not independently accessible: tuning a single-particle coupling generally modifies several multiparticle couplings simultaneously. Fractional revival between two chosen multiparticle configurations must therefore be achieved using only the restricted set of parameters available at the single-particle level. 
Fortunately, the inheritance of $\alpha$-scaled latent symmetry through a singlet provides the means to overcome this difficulty: Using graph-theoretical tools, we can identify a family of couplings that preserve the symmetry in the single-particle graph.
Due to inheritance, these couplings can therefore be varied without destroying the $\alpha$-scaled latent symmetry at the two-photon level.
By tuning these parameters, we can adjust the spectrum of the two-photon graph while retaining the structural constraint required for fractional revival. This provides a means of tuning the two-photon system to realise fractional revival between $\ket{\vec{u}}$ and $\ket{\vec{v}}$, without requiring independent control of the multiparticle couplings.

To illustrate this mechanism, consider the single-particle graph shown in \cref{fig:fractionalRevival}(a). For $u=2$ and $v=6$, this graph possesses an $\alpha=2$ scaled latent symmetry, with $s=4$ serving as the singlet.
In the corresponding two-boson graph, the symmetry relates the configurations $\ket{\vec{u}}=\ket{(2,4)}$ and $\ket{\vec{v}}=\ket{(4,6)}$.
As can  be easily shown, the symmetry is invariant under variation of the couplings $h_1,h_2,v_2,v_4$ [cf. \cref{fig:fractionalRevival}(a)].
After optimizing these parameters to satisfy the fractional-revival conditions, the evolution at $t_\mathrm{F}\approx 5.72$ yields
\begin{equation*}
    \hat{\mathcal{U}}(t_\mathrm{F})\ket{\vec{u}}
    \approx
    \left(-0.435+0.449\,\mathrm{i}\right)\ket{\vec{u}}
    +
    \left(0.706-0.333\,\mathrm{i}\right)\ket{\vec{v}}.
\end{equation*}
This fulfils the normalization condition and \cref{eq:realPartCond}, up to a negligible absolute error of $10^{-5}$.
This example demonstrates that fractional revival between multiparticle configurations can be achieved by tuning only the underlying single-particle graph.
The construction is not completely unrestricted: the accessible amplitudes are constrained by $\alpha$, the singlet condition, and the spectrum of the underlying graph. Systematically identifying which multiparticle superpositions can be generated in this way, and how accurately they can be realised, remains an interesting question for future work.

\prlsec{Quantum induced symmetries}%
So far, we have examined how a latent or geometric symmetry already present in the single-particle system carries over to a multiparticle system. 
We now demonstrate that the multiparticle condition (\ref{eq:multi-partLSMain}) can hold even when the single-particle condition (\ref{eq:paritySinglePart1}) is not satisfied. This is possible because each of the multiparticle eigenmodes corresponds to one possible distribution of the $K$ particles over the single-particle eigenmodes. Each amplitude of the multiparticle eigenmodes is obtained by combinations of several amplitudes of single-particle eigenmodes, captured by taking the permanent (bosons) or determinant (fermions) of selected rows and columns of the single-particle eigenmode matrix (End Matter).
This mixing of single-particle eigenmode amplitudes through the permanents (determinants) leads to the intriguing result that latent symmetries can be induced by quantum indistinguishability in inherently asymmetric systems.

An example of this quantum-induced latent symmetry is shown in \cref{fig:emergSymm}. The depicted graph possesses neither spatial nor latent symmetries at the single-particle level. 
Yet, for two particles, distinct pairs of latent symmetric sites emerge for bosons as well as fermions solely via of the combinatorics of the two-particle space. 
In other words, a multiparticle system can display latent symmetry even though the underlying physical system has no symmetry whatsoever -- the symmetry is induced purely by quantum indistinguishability.

To gain insight into how common this phenomenon would be, we carried out an exhaustive search over all uniform graphs of up to $12$ sites, looking for quantum-induced latent symmetry; cf.~\cref{tab:numberOfGraphs}.
This effect strongly depends on the nature of the particles: at $K=2$ we find only 7 bosonic graphs with this property, compared to 617 fermionic ones up to $N=11$ and additionally 14080 for $N=12$. We moreover examined all 8-, 9- and 10-site graphs for $K=3,4$ particles. For 3 particles, the exhaustive search returns only 5 graphs satisfying the condition for fermions -- most of the symmetries that emerged at $K=2$ disappear. For 4 particles, we found 579 fermionic graphs with a latent symmetry that did not have any symmetry for $K<4$; no bosonic graph with this property was found.
All graphs with quantum-induced latent symmetry in the multiparticle level can be found in a Supplemental File.

\begin{table}
    \centering
    \begin{adjustbox}{max width=\linewidth}
    \begin{tabular}{cccccccccc}
        \toprule
        $N$ & total & GS & LS & 2F & 2B & 3F & 3B & 4F & 4B \\
        \midrule
        8  & 1117       & 7565          & 78       & 14  & 1 & 0 & 0 & 15 & 0 \\
        9  & 261080      & 129628     & 2247     & 9   & 0 & 0 & 0 & 55 & 0\\
        10 & 11716571 & 3876175 & 78489 & 348 & 1 & 5 & 0 & 509 & 0 \\
        11 & 1006700565 & 	209176185 & 3714397 & 246 & 0 & -- & -- & -- & --\\
        12 & 164059830476 & 20734232912 & 269642745 & 14080 & 6 & -- & -- & -- & -- \\
        \bottomrule
    \end{tabular}
    \end{adjustbox}
    \caption{Number of graphs of up to $N=12$ sites with emerging symmetries. For each size, \textit{total} gives the number of distinct uniform graphs. Out of these, \textit{GS} and \textit{LS} denote the number of graphs that feature a single-particle geometric and latent symmetry. \textit{2F/B}, \textit{3F/B}, \textit{4F/B} show the number of graphs where a latent symmetry only emerges for $K=2,3,4$ fermionic/bosonic particles.} 
    \label{tab:numberOfGraphs}
    
    \end{table}

\prlsec{Conclusion}%
In this work, we use graph-theoretical tools to show how geometric and latent symmetries in quantum systems transfer from the single-particle to the multiparticle regime, for fermions and bosons. 
Moreover, we show that it is possible to induce symmetries in completely asymmetric systems upon the introduction of multiple particles, purely through quantum indistinguishability.
Finally, by extending the analysis to scaled latent symmetries, we have introduced fractional revival as a mechanism for bosonic multiparticle-state generation. In particular, we have identified symmetry-preserving parameters of the single-particle graph that can be tuned without breaking the inherited $\alpha$-scaled latent symmetry of the two-boson graph. This enables fractional revival between selected two-boson configurations using only single-particle couplings, which are experimentally accessible, without requiring independent control of couplings in the multiparticle graph.\\
These results pave the way towards the design of quantum networks that enable efficient state transfer or generation only for multiparticle states, of a specific particle number, with possible applications in the generation of cluster states and entanglement distribution.\\
An interesting question concerns the origin of the quantum-induced latent symmetries and whether they might be traced back to a generalized kind of hidden symmetries.
This could help to unveil the origin of the imbalance between fermions and bosons. A logical next step is expanding the search to graphs with larger number of modes, with nonuniform coupling, as well as more particles.\\
It will be exciting to explore latent symmetries in multiparticle Hamiltonians that contain nonlinear or squeezing terms. An excellent question would be under which conditions latent symmetries may be induced in interacting many-body quantum systems.
With the capability of inducing, modifying, and suppressing symmetries as a function of particle number via quantum indistinguishability, latent symmetries offer new ways to shape the multiparticle quantum behavior of light and matter.

\section*{Author contributions}
J.H., M.R., and T.A.W.W. conceived this work.
J.H. and M.R. performed most of the calculations, with additional input from T.A.W.W. and T.G.
J.H. wrote most of the manuscript, with additional input from M.R. and T.A.W.W.
A.S,  M.H., and M.R. funded the project.
T.A.W.W. and A.S. supervised.
All authors discussed the results, edited the manuscript and contributed to the work. 
\section*{Acknowledgments}
We thank G.E. Sommer and  K. Weiher for fruitful discussions.
This work was financially supported by the National Natural Science Foundation of China (NSFC), grant number W2533029.
M.R. acknowledges financial support in the framework of the ``Eastern Institute for Advanced Study Postdoctoral Excellent Programme''.
A.S. acknowledges funding from the Deutsche Forschungsgemeinschaft (grants SZ 276/27-1, SZ 276/34-1, and GRK 2676/1-2023 ‘Imaging of Quantum Systems’, project no. 437567992) and the Krupp von Bohlen and Halbach Foundation. A.S. and M.H. acknowledge funding from the Deutsche Forschungsgemeinschaft (SFB 1477 ‘Light–Matter Interactions at Interfaces’, project no. 441234705).

\cleardoublepage
\section{End Matter}
\appendix 
\prlsec{Deriving the isospectral reduction}%
We partition the system into two parts, $S$ and its complement $\overline{S}$, and write the eigenvalue equation $H \vec{\Phi} = E \vec{\Phi}$ in block form as
\begin{equation*}
    \begin{pmatrix}
        H_{SS} & H_{S\overline{S}}\\
        H_{\overline{S}S} & H_{\overline{SS}}
    \end{pmatrix}
    \begin{pmatrix}
        \Phi_{S} \\
        \Phi_{\overline{S}}
    \end{pmatrix} = E     \begin{pmatrix}
        \Phi_{S} \\
        \Phi_{\overline{S}}
    \end{pmatrix} \,.
\end{equation*}
Eliminating $\Phi_{\overline{S}}$, that is, the $\overline{S}$ components of $\vec{\Phi}$, we obtain the non-linear eigenvalue problem
\begin{equation} \label{eq:nlEVP}
    \mathcal{R}_S[H,E] \Phi_S = E \Phi_S\,,
\end{equation}
with
\begin{equation}
    \mathcal{R}_S[H,E]\equiv H_{SS} + H_{S\overline{S}} \left(E-H_{\overline{SS}} \right)^{-1} H_{\overline{S}S}
\end{equation}
the ISR of $H$ over $S$.
We remark that, under very mild conditions, the spectrum of $\mathcal{R}_S[H,E]$ coincides with that of the full Hamiltonian $H$ \cite{Bunimovich2014IsospectralTransformationsNewApproach}.

\prlsec{Construction of the multiparticle graph}
The construction is illustrated in Fig.~\ref{fig:fockGraphIntro}. Each site of the multiparticle graph corresponds to one possible distribution of the $K$ particles over the $N$ modes. For distinguishable particles, this distribution is an ordered assignment $\boldsymbol{\sigma}=(\sigma_1,\dots,\sigma_K)$, with $\sigma_k$ the mode occupied by particle $k$, so that the number of sites rises to $N^K$. Exciting modes 1 and 2 then populates either site $(1,2)$ or $(2,1)$, depending on the particle order. The graph is given by $H_\mathrm{Dist}=H\oplus\dots\oplus H=H^{\oplus K}$, with $\oplus$ the Kronecker sum.

For indistinguishable particles, on the other hand, the sites $(1,2)$ and $(2,1)$ describe the same physical state and merge into a single occupation configuration $\vec{m}=(m_1,\dots,m_N)$, where $m_n$ is the number of particles in mode $n$ and $\sum_n m_n=K$, strongly reducing the number of sites (Fig.~\ref{fig:fockGraphIntro}(b)). For indistinguishable fermions, the configurations corresponding to two particles on the same mode are additionally forbidden by the Pauli exclusion principle. This yields $\binom{N+K-1}{K}$ sites for bosons and $\binom{N}{K}$ for fermions.

The edges of the multiparticle graph correspond to an allowed hopping in the single-particle picture, one particle at a time, and are given by the matrix elements of the second-quantized Hamiltonian $\ham=\sum_{i,j}H_{i,j}\hat{a}_i^\dagger \hat{a}_j$ in the occupation basis. A hop moving one particle from mode $j$ to mode $i\neq j$ connects the corresponding configurations $\vec{m}$ and $\vec{m}'$ with weight $\langle\vec{m}'|\ham|\vec{m}\rangle$. The particle type is included through the (anti)commutation relations of the respective creation and annihilation operators. For bosons, we get $\langle\vec{m}'|\ham|\vec{m}\rangle=H_{i,j}\sqrt{m_j(m_i+1)}$, where $m_i,m_j$ are the occupations of the initial configuration $\vec{m}$. For $K=2$ bosons this gives a factor of $\sqrt{2}$ for edges incident to a site corresponding to both particles in the same mode (Fig.~\ref{fig:fockGraphIntro}(b), right). For fermions the occupations are restricted to $m_n\in\{0,1\}$ and the weight reduces to $\pm H_{i,j}$, with a sign flip whenever the number of occupied modes between $i$ and $j$ is odd, arising from the anticommutation relations (Fig.~\ref{fig:fockGraphIntro}(b), left).

\prlsec{$K$-particle latent symmetry condition}%
The latent symmetry condition on the multiparticle graph can be expressed entirely through the eigenvectors of the single-particle graph, without constructing the multiparticle graph.
For $K$-particle dynamics, we use the second-quantized Hamiltonian $\ham$ defined above, with $\hat{a}_i^\dagger$ ($\hat{a}_i$) the creation (annihilation) operator of a boson or fermion in mode $i$.

The eigenvalues and eigenmodes of the single-particle Hamiltonian $H$ are given by $H V=V\Lambda$, where the system's eigenmodes $\ket{\lambda_r}$ are the columns of the unitary matrix $V$, such that $\braket{i|\lambda_r}=V_{ir}$, and the entries of the diagonal matrix $\Lambda$ are its eigenvalues $\lambda_r$. Since $H$ is real and symmetric, $V$ can be chosen real.

We define the mode occupation configuration as $\vec{m}=(m_1,\dots,m_N)$, with $\sum_im_i=K$ (and $m_i\in\{0,1\}$ for fermions).
In the eigenmode basis the creation operator $\hat{b}^\dagger_r=\sum_iV_{ir}\hat{a}_i^\dagger$ creates a particle in eigenmode $r$: $\hat{b}_r^\dagger\ket{0}=\ket{\lambda_r}$. 
The eigenmode occupation $\vec{\Phi}=(\Phi_1,\dots,\Phi_N)$ then states how many particles occupy each single-particle eigenmode $\ket{\lambda_r}$ with $\sum_r\Phi_r=K$ (and again $\Phi_r\in\{0,1\}$ for fermions).

The mode-occupation basis state $\ket{\vec{m}}$ and the eigenmode-occupation basis state $\ket{\vec{\Phi}}$, i.e.\ the eigenstates of the multiparticle graph, are defined as
\begin{align}
    \ket{\vec{m}}&=\prod_i\frac{(\hat{a}_i^\dagger)^{m_i}}{\sqrt{m_i!}}\ket{0},\\
    \ket{\vec{\Phi}}&=\prod_r\frac{(\hat{b}_r^\dagger)^{\Phi_r}}{\sqrt{\Phi_r!}}\ket{0}.
\end{align}
For bosons, expanding the eigenmode operator in the mode basis, their overlap is given by the permanent of the submatrix of $V$ selected by $\vec{m}$ and $\vec{\Phi}$ \cite{scheel2008}
\begin{align}
    \braket{\vec{\Phi}|\vec{m}}_\mathrm{B}&=\left(\prod_im_i!\prod_r\Phi_r!\right)^{-1/2}\mathrm{perm}\,V\,[\vec{m}|\vec{\Phi}],
\end{align}
where $V[\vec{m}|\vec{\Phi}]$ is the $K\times K$ matrix whose rows are the mode indices of the $K$ particles (mode $i$ appearing $m_i$ times) and whose columns are their eigenmode indices (eigenmode $r$ appearing $\Phi_r$ times), with entries $V_{ir}$.

For fermions, the anticommuting operators replace the permanent by a determinant, such that
\begin{align}
    \braket{\vec{\Phi}|\vec{m}}_\mathrm{F}&=\mathrm{det}\,V\,[\vec{m}|\vec{\Phi}].
\end{align} 

In the absence of degeneracies in the $K$-particle spectrum, two basis states $\vec{m}$ and $\vec{n}$ are latent-symmetric if the overlap for both states has definite parity,
\begin{align}
    \braket{\vec{\Phi}|\vec{m}}_{\mathrm{B/F}}=\pm\braket{\vec{\Phi}|\vec{n}}_{\mathrm{B/F}},\ \forall \vec{\Phi},
    \label{eq:multi-partLS2}
\end{align} 
and there is no spatial symmetry between the two states.
Since the overlaps are permanents (determinants) of submatrices of $V$, i.e.\ sums of products of single-particle eigenvector components, Eq.~\eqref{eq:multi-partLS2} can be fulfilled for a pair of basis states $\vec{m},\vec{n}$ even if no pair of modes $i,j$ satisfies $V_{ir}=\pm V_{jr}$ for all $r$, i.e.\ if the single-particle graph has neither a spatial nor a latent symmetry. In such fully asymmetric systems, a symmetry emerges only for $K\geq2$ bosons or fermions, which we refer to as quantum-induced.

\prlsec{State generation}%
The optimized values for $v_2$, $h_1$, $v_4$, $h_2$ are $-0.0456799$, $0.284385, -1.12826$, $-0.446722$, $5.71878$. The spectral condition in \cref{fig:fractionalRevival}(b) is computed as
\begin{equation}
\left| \Psi_u + 
 \frac{\Psi_u^* \Psi_v}{\Psi_v^*} - \left(1/\alpha - \alpha \right) \Psi_v \right|^2 \,,
\end{equation}
which follows from \cref{eq:realPartCond}, and where $\Psi_{u,v}$ denote the components of the time-evolved wave function at sites $u,v$, respectively.

\prlsec{Symmetry inheritance}
Let modes $u,v$ be latently symmetric in the single-particle Hamiltonian
$\ham^{(1)}=\sum_{ij}H_{ij}\ket{i}\bra{j}$.
Then there exists a real-symmetric, orthogonal matrix
$q^{(1)} := \sum_{ij} q_{ij}\ket{i}\bra{j}$ that commutes with $\ham^{(1)}$ and
exchanges modes $u,v$
\cite{Rontgen2021AQSymmetriesMatrixItsIsospectral,Godsil2024AJC881StronglyCospectralVertices,Rontgen2021PRL126180601LatentSymmetryInducedDegeneracies}.
It can be constructed as $P_+ - P_-$, where $P_{+/-}$ are the projectors onto the
eigenmodes with positive and negative parity on $u,v$, respectively.

The matrix $q^{(1)}$ is block-diagonal.
The first block corresponds to the union of the sites $u,v$ (where it acts as the
exchange operator) and all singlet sites (if any), where it acts as the identity.
We call this set of sites $S$.
The second block corresponds to all other sites $\overline{S}$; in general, this
block is a full orthogonal matrix.

We now construct an operator $\Gamma$ that acts on the $K$-particle sector as 
$q^{(1)}\otimes\dots\otimes q^{(1)}$ ($K$ factors).
We then have $\Gamma \ket{0} = \ket{0}$, and also
\begin{equation}
    \Gamma\, \hat{a}_i^\dagger\, \Gamma^{-1} = \sum_j q_{ji}\, \hat{a}_j^\dagger .
\end{equation}

Because $q^{(1)}$ is real, orthogonal and symmetric, $\Gamma$ is real, unitary and Hermitian.
Most importantly, $\Gamma$ commutes with the second-quantized Hamiltonian $\ham=\sum_{ij}H_{ij}\hat a_i^\dagger \hat a_j$.
Consider now the two $K$-particle bosonic Fock states
\begin{equation} \label{eq:latentFockStates1}
    \ket{K,0}_{u,v} =\frac{(\hat{a}_u^{\dagger})^K}{\sqrt{K!}}\ket{0},
\qquad
\ket{0,K}_{u,v} =\frac{(\hat{a}_v^{\dagger})^K}{\sqrt{K!}}\ket{0},
\end{equation}
that live exclusively on mode $u$ or $v$, respectively.
Since $q_{ui}$ ($q_{vi}$) vanishes unless $i=v$ ($i=u$), we have
$\Gamma\hat{a}_u^\dagger\Gamma^{-1}=\hat{a}_v^\dagger$ and
$\Gamma\hat{a}_v^\dagger\Gamma^{-1}=\hat{a}_u^\dagger$, and hence
$\Gamma\ket{K,0}_{u,v}=\ket{0,K}_{u,v}$ and $\Gamma\ket{0,K}_{u,v}=\ket{K,0}_{u,v}$.
These two states therefore span an invariant subspace of $\Gamma$, on which
$\Gamma$ acts as the exchange matrix $P=\begin{psmallmatrix}0&1\\1&0\end{psmallmatrix}$.
As $\Gamma$ is Hermitian, the orthogonal complement is invariant as well, so
$\Gamma$ is block-diagonal with respect to this subspace.
Because $\Gamma$ commutes with $\ham$, it follows that the isospectral reduction of
$\ham$ over $\ket{K,0}_{u,v},\ket{0,K}_{u,v}$ commutes with $P$
\cite{Rontgen2021AQSymmetriesMatrixItsIsospectral}; these two states are thus
latently symmetric.
We remark that, for fermions, $\ket{K,0}_{u,v}$ does not exist for $K>1$.

Next, let $s_1,\ldots,s_{n}$ (with $n = K-m$) be (possibly non-distinct) singlets, that is, sites on which
$q^{(1)}$ acts as the identity. Then
$\Gamma \hat{a}_{s_j}^\dagger\Gamma^{-1}=\hat{a}_{s_j}^\dagger$, and the two Fock states
\begin{align} \label{eq:latentFockStates2}
    (\hat{a}_u^{\dagger})^m \hat{a}_{s_1}^{\dagger} \ldots \hat{a}_{s_n}^{\dagger}\ket{0}, \quad
    (\hat{a}_v^{\dagger})^m \hat{a}_{s_1}^{\dagger} \ldots \hat{a}_{s_n}^{\dagger}\ket{0}
\end{align}
are exchanged by $\Gamma$.
Thus they form an invariant subspace of $\Gamma$ on which it acts as $P$, and by the
same argument as above, these two states are latently symmetric.

Let us now restrict to $m=1$.
For $\alpha$-scaled latent symmetries, $q^{(1)}$ is still real-symmetric, block-diagonal and 
commutes with $\ham^{(1)}$, but it is no longer orthogonal: on the sites $u,v$ it
acts as
$$
\begin{pmatrix}0&1\\1&\beta\end{pmatrix},\qquad \beta=\alpha-\frac{1}{\alpha}\neq0,
$$
while it acts on a singlet $s$ as $x$-times the identity operation, with some scalar $x$.
Then
$$
\Gamma\hat a_u^\dagger\Gamma^{-1}=\hat a_v^\dagger,\qquad
\Gamma\hat a_v^\dagger\Gamma^{-1}=\hat a_u^\dagger+\beta\hat a_v^\dagger,\qquad
\Gamma\hat a_{s_j}^\dagger\Gamma^{-1}=x \hat a_{s_j}^\dagger .
$$
Consequently, the two Fock states \cref{eq:latentFockStates2} span an invariant
subspace of $\Gamma$ on which it acts as
$$
X=\begin{pmatrix}0&1\\1&\beta\end{pmatrix}.
$$
Hence, if $u,v$ are $\alpha$-scaled latently symmetric and $s_1,\ldots,s_n$ are
singlets, the states \cref{eq:latentFockStates2} are $\alpha$-scaled latently symmetric.

\begin{figure}[hb!]
    \centering
    \includegraphics[width=0.85\linewidth]{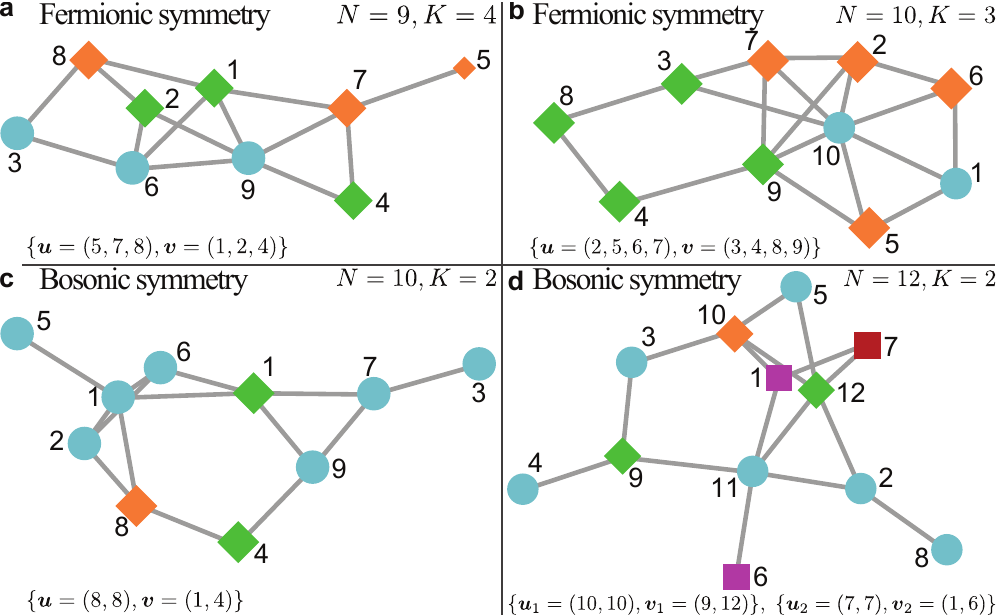}
    \caption{Four example graphs without any conventional symmetry and no latent symmetry in the coupling graph. \textbf{a} 9-site system, symmetry for $4$ fermions, \textbf{b} 10-site system, symmetry for 3 fermions \textbf{c} 10-site, symmetry for 2 bosons \textbf{d} 12-site system, symmetry for 2 bosons}
    \label{fig:SMGraphs}
\end{figure}

\prlsec{More example graphs}
In \cref{fig:SMGraphs} four more example graphs with quantum induced latent symmetry are displayed.

\end{document}